\documentclass[aps,prl,twocolumn,groupedaddress]{revtex4}
\usepackage{graphicx,amsmath,amssymb,natbib}
\usepackage{color}
\usepackage{multirow}
\begin{document}
\bibliographystyle{h-physrev3}

\title{Going beyond the double well: complex mode dynamics of effective 
coupled oscillators in infinite dimensional systems}
\author{T. J. Alexander$^{1}$}\email{t.alexander@unsw.edu.au} \author{D. Yan$^2$} \author{P. G. Kevrekidis$^2$} 

\affiliation{$^1$School of Physical, Environmental and Mathematical Sciences, University of New South Wales at the Australian Defence Force Academy, Canberra, Australia 2600}
\affiliation{$^2$Department of Mathematics and Statistics, University of Massachusetts, Amherst, Massachusetts 01003-4515, USA}

\begin{abstract}
In this work we explore how nonlinear modes described by 
a dispersive wave equation (in our example, the nonlinear Schr\"{o}dinger equation) and localized in a few wells of a periodic potential can act analogously to a chain of coupled mechanical oscillators.  We identify the small-amplitude oscillation modes of these `coupled wave oscillators' and find that they can be extended into the large amplitude regime, where some can `ring' for long times.  We also identify prototypical case examples of more complex dynamical behaviour
that can arise in such systems 
including the breakdown of Josephson-like oscillations and of internal 
modes more generally, 
the transfer of energy out of/destabilization of fundamental oscillation 
modes and the emergence of chaotic oscillations for large amplitude 
excitations.  
We provide details of the phase perturbations required for experimental 
observations of such dynamics, and show that the oscillator formalism can be 
extended to predict large amplitude excitations in genuinely two-dimensional 
configurations.
\end{abstract}

\pacs{03.75.Lm, 03.75.Kk, 05.45.Yv}

\maketitle

Since Huygens first noticed the synchronization of two pendula attached to the same wall \cite{BennettPRSLA2002}, the coupling of mechanical oscillators has fascinated scientists and the general public alike.  Complex and often counter-intuitive results are found in even simple coupled systems such as the spring-pendulum \cite{OlssonAJP1976} and two coupled pendula \cite{DenardoAJP1999}.  With the advent of exotic coherent wave systems the oscillator formalism has found new life, providing a deep understanding of effects such as the Josephson tunnelling of supercurrents \cite{JosephsonAP1965}, the evanescent wave tunnelling between two optical waveguides \cite{SnyderJOSA1972} and Josephson-like oscillations in a Bose-Einstein condensate (BEC) confined in a double-well potential \cite{JavanainenPRL1986,AlbiezPRL2005}.  Often, the oscillatory dynamics in these coherent wave systems are captured by the dynamics of a single mechanical oscillator \cite{JosephsonAP1965,SmerziPRL1997}.  In contrast, we explore here how 
matter waves or optical waves in a few wells of a periodic potential, such as light in coupled optically-induced waveguides \cite{ChristodoulidesNat2003}, or a BEC in an optical lattice \cite{BlochNP2005}, may behave analogously to coupled mechanical oscillators.  By focusing on a few (more than one) 
degrees of freedom, 
we predict new macroscopic wave dynamics, and so open up the possibility of 
experimental investigation of coupled effective oscillator dynamics in a 
precisely controlled environment.

Tunnelling between multiple wells has been observed in BEC in a periodic potential \cite{CataliottiSci2001}, however our work differs by beginning with a stationary configuration of occupied sites in a periodic potential \cite{AlexanderPRL2006,AlexanderPRA2011}, and examining the tunnelling dynamics which may occur between these nonzero (``excited'') sites.  Rather than trying to characterize all the possible dynamics, as has been the emphasis in double-well \cite{RaghavanPRA1999} and triple-well \cite{LiuPRA2007,kapkev} studies, we focus on the periodic excitations, analogous to the modes of a coupled-oscillator system.  This allows us to go beyond a triple-well configuration and identify the amplitude/phase oscillation modes which may be excited by appropriate initial conditions
for multi-site excitations.  We also explore a number of instability pathways 
possible in the multi-well configurations and extend the oscillator 
formalism from a one-dimensional (finite) 
chain to a truly two-dimensional arrangement.  We note that a recently published work \cite{SacchettiPD2012} complements our results with the detailed derivation of a model for the $N$-well case and an examination of the stationary states in four coupled wells.

\begin{figure}
\includegraphics[width=1.0\columnwidth]{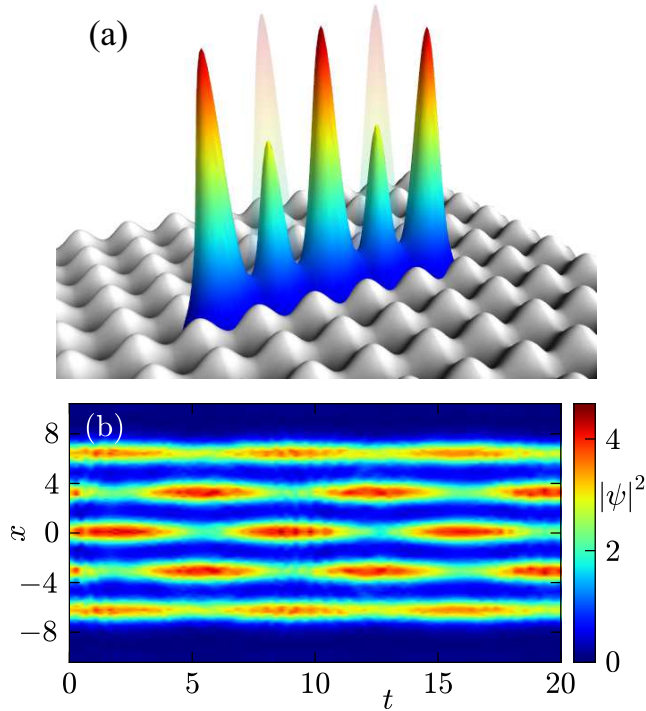}
\caption{(color online). (a) Schematic of a five-site state localized in a periodic potential, showing large scale periodic excitation; (b) Appearance of regular nonlinear wave oscillations  following initial phase perturbation.  The
colorbar on right shows density/intensity according to color.  Parameters: $\mu = 6$, $V_0 = 6$, $\epsilon = \pi/4$.}
\label{fig1}
\end{figure}

As our prototypical case example, we use systems described by the two-dimensional nonlinear Schr\"{o}dinger equation with a repulsive nonlinearity,
\begin{equation}
i\frac{\partial \Psi}{\partial t}+\frac{\partial^2 \Psi}{\partial x^2} + \frac{\partial^2 \Psi}{\partial y^2} - V(x,y)\Psi - |\Psi|^2\Psi = 0, \label{NLS}
\end{equation}
and a spatially periodic external potential of depth $V_0$ given by $V(x,y) = V_0(\sin^2(x)+\sin^2(y))$.  This equation appears in different physical contexts, with the most relevant that of a Bose-Einstein condensate (BEC) in an optical lattice \cite{MorschRMP2006}(see e.g. \cite{AlexanderPRA2011} for an example of a conversion between the normalised equation (\ref{NLS}) and physical parameters in BEC), or a coherent light field in a defocusing material with an optically induced periodic potential \cite{ChristodoulidesNat2003}.  We are interested in the possibility of coupled coherent wave oscillators described by this model, and as a basis for this study we examine the truncated nonlinear Bloch waves supported by the periodic potential \cite{AlexanderPRL2006}.  These states exist inside the linear band-gap of the periodic potential through a balance between the nonlinearity and Bragg reflection, and form distinct families depending on the number of occupied sites \cite{WangPRA2009}.  In the two-dimensional potential the configuration of the occupied sites can assume complex 
forms through suitable contours \cite{AlexanderPRA2011}, however for the purposes of this work we focus primarily on a finite chain of occupied sites.  In Fig. \ref{fig1}(a) we show an example of a five site state undergoing the large amplitude oscillatory dynamics of interest to this work.  With an appropriate phase perturbation applied to the occupied sites the initially almost equal 
amplitude state starts to periodically exchange particles in a coherent and long-lived oscillation (shown in Fig. \ref{fig1}(b)).   While this 
dynamics is shown for the model (\ref{NLS}) we note that the primary results of our work are independent of the specific physical details of the problem, such as the form of the potential or the precise nature
of the underlying infinite dimensional wave equation.

To better reveal the generality of our results we begin by considering the discrete version of model (\ref{NLS}) obtained from the ansatz $\Psi = \sum_i^M \sqrt{N_i(t)}\exp(i(\theta_i(t)-\mu t))\phi_i(x,y)$, where $\phi_i(x,y)$ captures the spatial dependence of the wavefunction at site $i$ for a stationary state with $M$ occupied sites and chemical potential $\mu$.  Substitution of this ansatz into (\ref{NLS}) and integration over the spatial dimensions leads to time-dependent equations for $N_i$ and $\theta_i$, the particle number and phase respectively of site $i$.  See \cite{RaghavanPRA1999} and \cite{LiuPRA2007} for the derivation in the two-site and three-site cases respectively; a derivation for the general $M$ site case has recently appeared in~\cite{SacchettiPD2012}. Moreover, such models can be considered as finite dimensional truncations
of the infinite chain cases examined e.g. in~\cite{SmerziPRA2003,AlfimovPRE2002}.
 However, unlike these earlier works, we consider also the effect of the next order nonlinear coupling terms using a nonlinear basis, applicable for the strong coupling between sites seen in our continuum stationary states.  The resulting equations can be found from the Hamiltonian,
\begin{align}
H = &\sum_{i=1}^{M-1} \left[2(K_{i,i+1}-\eta_{i,i+1}(N_i+N_{i+1}))\sqrt{N_iN_{i+1}}\right.\\\nonumber
&\left.\cos(\theta_{i+1}-\theta_i) - \chi_{i,i+1} N_iN_{i+1} \left(2+\right.\right.\\\nonumber
&\left.\left.\cos(2(\theta_{i+1}-\theta_i))\right)\right] -\sum_{i=1}^{M}\frac{\gamma_i}{2}N_i^2 +E_iN_i,
\label{hamil}
\end{align}
using Hamilton's equations for the canonical coordinates $(\theta_i,N_i)$,
\begin{equation}
\frac{d\theta_i}{dt} = \frac{\partial H}{\partial N_i}, \;\; \frac{dN_i}{dt} = -\frac{\partial H}{\partial \theta_i}, \;\; i = 1..M-1.  
\label{system}
\end{equation}
As the total mass is conserved, $\sum_i^M N_i = N_T$, we may set $N_M = N_T - \sum_i^{M-1} N_i$, leaving $M-1$ free mass variables.  Similarly we have only $M-1$ free phase variables, defined by the phase differences $(\theta_{i+1} - \theta_i)$.  The system (\ref{system}) is thus equivalent to a mechanical system with $M-1$ degrees of freedom, or more precisely a set of $M-1$ coupled momentum-shortened mechanical pendula with phase $(\theta_{i+1}-\theta_i)$ and conjugate momentum $N_i$, and higher-order nonlinear coupling given by the $\eta$ and $\chi$ terms.  The system coupling parameters are given by $K_{i,i+1} = -\int_{-\infty}^{\infty} \nabla \phi_i \nabla \phi_{i+1} + \phi_i V({\bf r})\phi_{i+1} - \mu\phi_i\phi_{i+1}d{\bf r}/P_i$, $\eta_{i,i+1} = \int_{-\infty}^{\infty} \phi_i^3\phi_{i+1}d{\bf r}/P_i$ and $\chi_{i,i+1} = \int_{-\infty}^{\infty} \phi_i^2\phi_{i+1}^2d{\bf r}/P_i$, where $P_i = \int_{-\infty}^{\infty} \phi_i^2d{\bf r}$.  The effective on-site nonlinearity is given by $\gamma_i = \int_{-\infty}^{\infty} \phi_i^4 d{\bf r}/P_i$ and the on-site energy, equal between occupied sites in our case, 
is given by $E_i = \int_{-\infty}^{\infty} (\nabla \phi_i)^2 + \phi_i V({\bf r})\phi_i - \mu\phi_i^2 d{\bf r}/P_i$. Here the subscripts $i$ and $i+1$ refer
to modes localized in adjacent wells ${i,i+1}$ of the periodic potential.  
Hence, for instance, in order to obtain these parameters 
in the restricted two-mode case, we use $\phi_{1,2} = (\Psi_s\pm\Psi_a)/(2\sqrt{N_s})$ where $\Psi_s$ and $\Psi_a$ are the symmetric and anti-symmetric two-site solutions to the full system (\ref{NLS}) and $N_s = \int_{V_1} |(\Psi_s+\Psi_a)/2|^2d{\bf r}$ is the onsite mass obtained by integrating over a single occupied well $V_1$ of the periodic potential $V(x,y)$.  For the case of the continuum stationary state with $V_0 = 6$ and $\mu = 6$ the associated values are $K = 0.17337$, $\eta = 0.00567$, $\chi = 0.00085$, $\gamma = 0.196$, $N_0 = 7.64$ and $E = 1.519$ (the same for both sites).

\begin{figure}
\includegraphics[width=1.0\columnwidth]{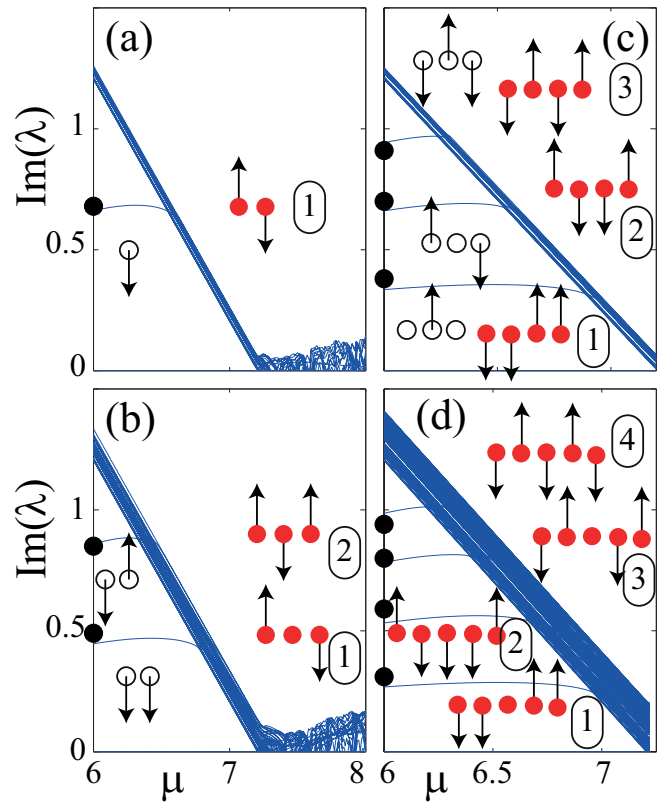}
\caption{(color online). Internal mode oscillation frequency versus chemical potential $\mu$ found from linear stability analysis for (a) two-site; (b) three-site; (c) four-site; and (d) five-site states.  Solid black circles show eigenvalues from the discrete system at $\mu = 6$.  Insets show relative change in density/power between sites for each internal mode (filled circles with arrows indicating direction of change) and analogous coupled effective pendulum modes (open circles).}
\label{fig2}
\end{figure}

The free boundary conditions of the oscillator chain suggest that 
the ground state occupied site configuration does not have equal particle numbers at each site (see the discussion for the three-site case \cite{LiuPRA2007}
and also the results of~\cite{SacchettiPD2012} for the four site one).  However, we are interested in the oscillation modes of the ground state rather than details of the ground state itself, so we make the approximation that $N_i = N_0$ for all sites (exact in the limit $K\rightarrow 0$) and that all parameters are the same as in the two-site case.  These approximations allows us to obtain explicit forms for the mode structures and frequencies in the discrete model.  To find these modes, and their associated frequencies, we 
substitute $N_i = N_0+\delta_i$, $\theta_i = \epsilon_i$ into the system equations (\ref{system}) and keep only the linear terms in $(\delta_i,\epsilon_i)$.  We find that the stationary state is neutrally stable,
as is expected for our defocusing nonlinearity; in the focusing case all
the norm would tend to focus to a single well rendering the above state
highly unstable with $M-1$ real eigenvalue pairs; in the defocusing case,
there are $M-1$ imaginary eigenvalue pairs (corresponding to eigenfrequencies
of oscillation). 
We list the mode eigenfrequencies $\omega_j$ in Table \ref{freq}, 
where $j = 1 \dots M-1$ runs from the lowest to highest frequency eigenmode.  The analytical expressions are lengthy in the general case, so for the most part we present only the case with $\eta=\chi=0$.  The calculation of the eigenvalues in the case $\eta,\chi \ne 0$ amounts to the diagonalization of a $2M\times 2M$ matrix, which for 2 to 5 sites is a straightforward task.  The eigenvalues for the state $\mu = 6$, $V_0 = 6$ are shown as solid black circles in Fig. \ref{fig2}. 
\begin{table}[h]
\centering
\begin{tabular}{|l|l|}
\hline
Two-site & $\omega_1 = \sqrt{4(K-2\eta N_0)\gamma N_0 +4(K-2\eta N_0)^2}$ \\
\hline
\multirow{2}{*}{Three-site} & $\omega_1 = \sqrt{2K\gamma N_0 + K^2}$ \\ & $\omega_2 = \sqrt{6K\gamma N_0 + 9K^2}$ \\
\hline
\multirow{3}{*}{Four-site} & $\omega_1 = \sqrt{(4-2\sqrt{2})K\gamma N_0 + (6-4\sqrt{2})K^2}$ \\ & $\omega_2 = \sqrt{4K\gamma N_0 + 4K^2}$ \\ & $\omega_3 = \sqrt{(4+2\sqrt{2})K\gamma N_0 + (6+4\sqrt{2})K^2}$ \\
\hline
\multirow{4}{*}{Five-site} & $\omega_1 = \sqrt{(3-\sqrt{5})K\gamma N_0 + \frac{(7-3\sqrt{5})}{2}K^2}$ \\ & $\omega_2 = \sqrt{(5-\sqrt{5})K\gamma N_0 + \frac{(15-5\sqrt{5})}{2}K^2}$\\ & $\omega_3 = \sqrt{(3+\sqrt{5})K\gamma N_0 + \frac{(7+3\sqrt{5})}{2}K^2}$ \\ & $\omega_4 = \sqrt{(5+\sqrt{5})K\gamma N_0 + \frac{(15+5\sqrt{5})}{2}K^2}$ \\
\hline
\end{tabular}
\caption{Approximate mode eigenfrequencies for multi-site stationary states in the weak coupling limit.}
\label{freq}
\end{table}

We return now to our original continuum model (\ref{NLS}) and perform a similar linear stability analysis to determine the eigenfrequencies and eigenmodes.  We see in Fig. \ref{fig2} for the cases of $M=2 \dots 5$ wells, 
a confirmation of the presence of the $M-1$ oscillation eigenfrequencies (numbered in ascending order of frequency), 
as found for the discrete case.  We see that the discrete model predictions are excellent in the two site case, and still very good with a larger number of sites despite the assumption of equal well populations.  This shows that even in the strongly coupled case the vast simplification of the discrete model is still 
meaningful.  Included also in Fig. \ref{fig2} is a schematic representation of the associated eigenmode profiles found 
in both the discrete and continuum cases (filled circles).  The arrows indicate the direction of amplitude change for one half of an oscillation.  In the other half the directions of the arrows is reversed.  The mode structure shows interesting parallels to the more familiar mechanical oscillator normal modes.  Noting that the symmetry of the phase oscillations mirrors the symmetry of the amplitude oscillations, and remembering that the phase of the equivalent mechanical oscillator is given by $(\theta_{i+1} - \theta_i)$, we can represent schematically the analogue mechanical mode associated with each amplitude/phase oscillation mode.  These are shown on Fig. \ref{fig2} as $M-1$ open circles for 
each mode, with the arrows now indicating the direction of oscillation of the 
{\it effective pendulum}. It is interesting to point out here that this
effective normal mode approach may be more broadly applicable. For instance,
the works of~\cite{dark_sol} suggest that the dynamics of dark solitary waves
as longitudinally vibrating particles in BECs share similar symmetries
and characteristic oscillation modes.   

\begin{figure}
\includegraphics[width=1.0\columnwidth]{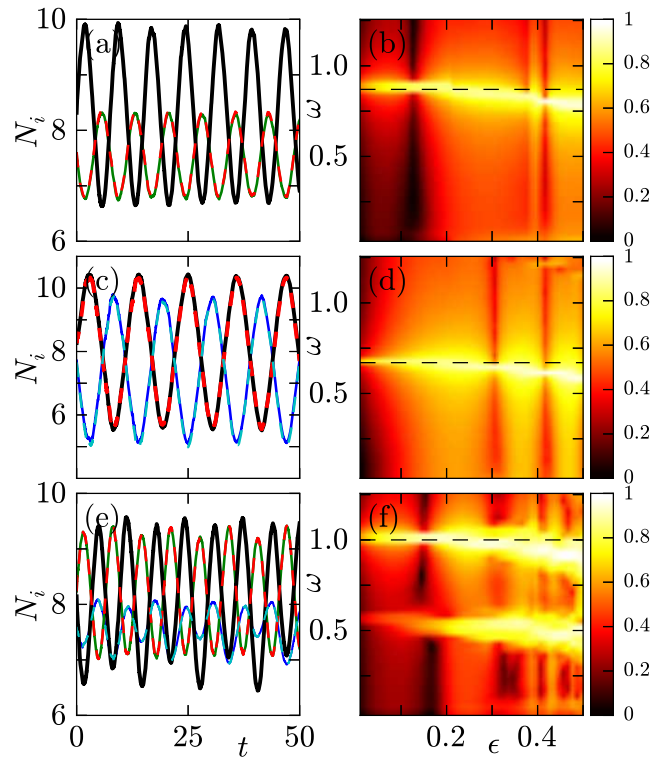}
\caption{(color online).  Mass/power oscillations per site for (a) three-site state, $\epsilon = 0.2\pi$ (thick solid line: central site); (c) four-site state, $\epsilon = 0.5\pi$ (solid lines: central sites); (e) five-site state, $\epsilon = 0.2\pi$ (thick black line: central site; thick dashed lines: adjoining sites; thin dashed lines: outer sites).  (b,d,f) Frequency dependence of resultant oscillations associated with perturbation $\epsilon$ for three-, four- and five-site cases, showing deviation from linear predictions (dashed lines) at large amplitude.  Data presented is the Fourier transform of trajectories propagating to $t=200$.}
\label{fig3}
\end{figure}

An important consideration for the identified 
oscillation modes is their behaviour in the 
presence of nonlinearity.  In the discrete case the momentum-shortening and nonlinear coupling of the 
effective pendula introduces a complexity absent in similar mechanical oscillator systems.
A key question therefore pertains to whether these oscillation modes
persist in the full nonlinear wave dynamics (and hence beyond
the associated linearization).  And if so, how can they be excited and
what role do they play in the system's nonlinear dynamics~?  The full analysis of these modes in the nonlinear case is beyond the scope of this work.
Here, we restrict ourselves to some vignettes showcasing the relevance
and potential observability of such modes and the intriguing nonlinear
dynamics that they may exhibit.
Our approach to generation is to look for mode symmetries which are preserved in the presence of the full system nonlinearity.  For instance one obvious symmetry is a reflection about the mid-point of the oscillator chain.  Mode 2 in
the three-site case, mode 2 in the four-site case and modes 2 as well as 4 
in the five-site case all possess this reflection symmetry (while the
rest are anti-symmetric). This symmetry is not broken in the presence of nonlinear coupling between the sites.  
To achieve their dynamical generation we make use of a second property of the linear modes: the phase oscillation is half a cycle out of phase with the amplitude oscillation, i.e. the phase is at the extremum of an oscillation, while the amplitude is at the mid-point.  This means that with an appropriate choice of phase, the oscillations may be initiated from the stationary state.  
To simulate the possible experimental conditions we apply a fixed phase shift between maxima of the periodic potential, but extending to the boundaries of our simulation domain in the dimension perpendicular to the oscillator chain.  The phase perturbations for the three-site, four-site and five-site modes we excite are respectively:  $(0,\epsilon,0)$, $(0,\epsilon,\epsilon,0)$ and $(0,-\epsilon/2,\epsilon/2,-\epsilon/2,0)$.  The oscillations of the particle numbers in each well, resulting from these phase perturbations, are shown in Fig. \ref{fig3} for the three-site, four-site and five-site modes mentioned above.  Note, $N_i = \int_{V_i} |\psi|^2 d{\bf r}$, where $V_i$ signifies the spatial domain of the $i$-th occupied well of the periodic potential $V(x,y)$.  We see that in each case the oscillations are large in amplitude and regular, illustrating the {\it 
persistence} 
of the linear modes into the nonlinear regime.  We show in Fig. \ref{fig3}(b,d,f) the dependence of the oscillation frequency on the phase perturbation $\epsilon$ (and corresponding amplitude of oscillation), and see that as the amplitude increases the oscillation frequency begins to deviate from the linear oscillation frequency, as expected for a nonlinear system.  The spectrum also shows the appearance of a frequency subharmonic in the five site case, which increases with the perturbation, ultimately leading to a strong perturbation of the linear mode.

\begin{figure}
\includegraphics[width=1.0\columnwidth]{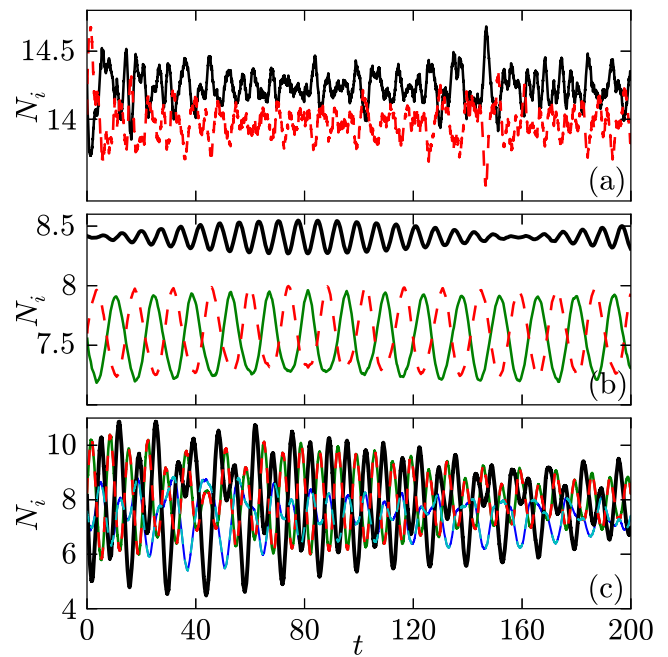}
\caption{(color online). Examples of instabilities: (a) Absence of Josephson-lik                  e oscillations for perturbations of the 
stationary state with $\mu = 7$ ($\epsilon = 0.1\pi$); (b) Unstable first linear mode in three-site state (transfer of energy to the 
central site with $\epsilon = \pm0.05\pi$ on outer sites); (c) Breakdown of mode 4 in the five-site case for $\epsilon = 0.4\pi$.  
The lines follow the same labelling convention as in Fig. \ref{fig4}. $\mu = 6$ in (b,c).}
\label{fig4}
\end{figure}

As may perhaps be expected, not all of the linear modes are robust in the 
highly nonlinear regime.
We focus here on some of the key instability/persistence failure
examples which may be observed in this coherent wave oscillator system.  The first illustrated case example where such a mode does not persist 
has no parallel in analogue mechanical systems, and stems purely from
the extended nature of the underlying system. 
This nonlinearity-induced 
`cut-off' in possible mode excitation occurs for all modes, however we consider the simplest two occupied site case.  In a two-mode reduction, the Josephson-like oscillation between the two sites always exists for a small enough perturbation from the ground state configuration $(N_0,N_0)$.  As can be seen in Fig. \ref{fig2}(a) a significant effect of the extended periodic 
lattice is that this oscillation ceases to exist, even in the infinitesimal limit, for stationary states with large enough chemical potential $\mu$.  Thus even though the stationary two-site state is stable, no localized internal
mode of amplitude oscillations is possible in this regime of large
$\mu$.  As seen in Fig. \ref{fig4}(a) an appropriate phase perturbation $(\epsilon,0)$ does not lead to any coherent tunnelling dynamics between the sites.  
In this regime, it is evident that the simple two-site approximation
breaks down, although the linearization captures this feature.              

The second considered pathway is commonly encountered in mechanical systems, and consists of mode instability due to nonlinearity~\cite{OlssonAJP1976,DenardoAJP1999}.  As an example we consider the first mode of the three-site case, which is analogous to the in-phase mode of two coupled pendula.  
We find this mode to be unstable 
for our coherent wave oscillator case, with the
dynamics (see Fig. \ref{fig4}(b)) apparently pointing to an oscillatory
instability. This can be attributed to a parametric instability due
to the nonlinearity-induced coupling between the different modes in a
way similar to what happens in the mechanical system analyzed 
in~\cite{DenardoAJP1999}. Here, linearization still offers the guiding
principles of the relevant modes, but nonlinearity destabilizes (some of)
them, via mode coupling.

The final  pathway that we consider for the drastic modification of linear
dynamics under strong nonlinearity is the breakdown of regular oscillations at large amplitudes (i.e., the breakdown of linearized analysis altogether
as a guiding principle for the dynamics).  As we can see in Fig. \ref{fig4}(c), at large amplitude (corresponding to initial phase perturbation $\epsilon = 0.4\pi$) the excitation of mode 4 in the five-site state ceases to be regular, as we also saw in the spectral profile of Fig. \ref{fig3}(f).  Oscillations persist, however the initial regular oscillation breaks up and some energy is lost through tunnelling to other wells.
As an interesting aside, the reflection symmetry is still maintained throughout the irregular dynamics.  If we look back to our discrete equations we see that for dynamics satisfying this reflection symmetry, chaos is indeed 
only possible for five sites or more.  The four site (reflection symmetric)
case amounts to effectively two sites but with two conservation laws, 
 so no chaos is possible. 

\begin{figure}
\includegraphics[width=1.0\columnwidth]{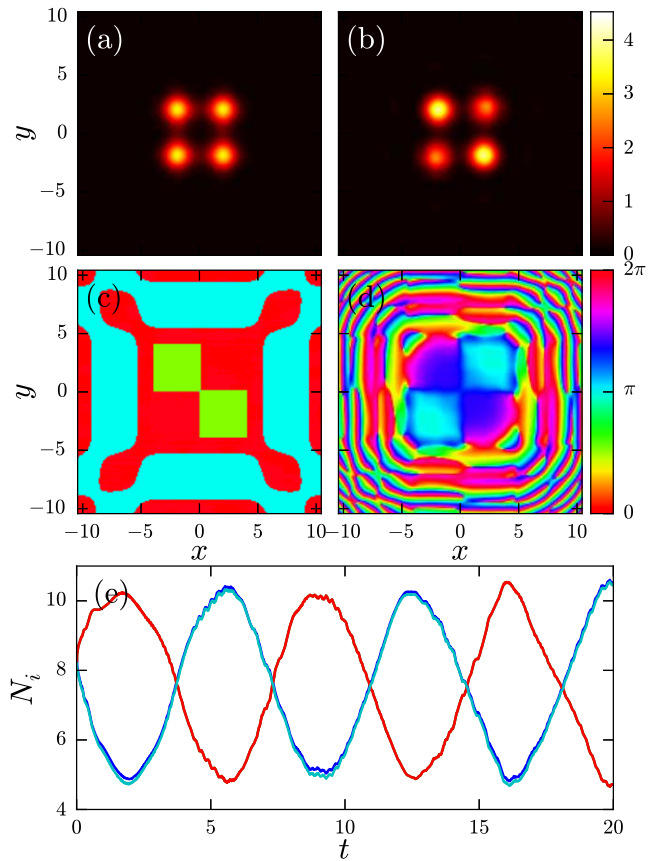}
\caption{(color online). Amplitude/phase oscillations for a two-dimensional four-site configuration. (a) Density and (c) associated phase with initial phase perturbation corresponding to $\epsilon = 0.5\pi$ at $t=0$ (b,d) corresponding density and phase at $t = 20$ (values given by color-bars on right). (e) Associated power-per-site oscillation.}
\label{fig5}
\end{figure}

 Finally, we show that the oscillator framework we have introduced may be easily extended to genuinely higher dimensional lattice 
configurations.  As an example we consider the simplest genuinely two-dimensional configuration, the four-site square shown in Fig. \ref{fig5}(a).  Phase oscillations of an initial vortex phase have been previously predicted in this configuration~\cite{AlexanderPRL2004} (see also~\cite{book} for the excitations of the lattice contour). Here, we 
illustrate the amplitude oscillations that may also be observed in this 
geometry.  Following the symmetry arguments of the chain configuration we 
note that this system is symmetric when reflected through a line running 
through the centre and two sites, and also through a line running through the centre and bisecting the occupied sites.  We find modes satisfying one of 
these symmetries are stable and we consider here the excitation of the 
diagonal mode.  By imposing a phase shift on the two diagonally opposite 
sites (see Fig. \ref{fig5}(c)), we initiate a coherent tunnelling oscillation between the sites (see Fig. \ref{fig5}(e)).  These oscillations correspond to a diagonal oscillation of phase and amplitude (Fig. \ref{fig5}(b,d)) and persist even in the large amplitude limit.  
We have observed such features (e.g., long-lived periodic oscillations) in more extended square-like states, however a detailed 
discussion of the complex oscillation patterns possible therein
is beyond the scope of this work and will be analyzed elsewhere.

In conclusion, we have presented some prototypical features
within the rich potential of
 a framework for understanding and predicting complex dynamics of wave systems 
in a periodic potential.  We have shown that Josephson-like oscillations can 
still be observed, and thus do not require the spatial confinement imposed by a double-well potential.  We have revealed that the Josephson-like dynamics may be understood as an oscillation mode of the stationary two-site state of the potential.  We have generalised this formalism to a chain of $M$ coupled sites, revealing that such a chain has $M-1$ linear oscillation modes.  We have 
reduced the corresponding dynamics to a discrete dynamical system, showing 
that the observed oscillations may be understood as the normal modes of
this effective oscillator system.  We have shown that some of the modes are robust even in the large-amplitude, nonlinear case, and we have provided a set 
of guidelines for their experimental excitation.  
On the other hand, we have also proposed a series of 
non-persistence/instability pathways that may arise, ranging from the
disappearance of Josephson-like modes, to instabilities due to mode coupling
and even to complex irregular dynamics beyond the mode description.
We have shown that this 
formalism can be extended to higher dimensions, and provides 
again a valuable theoretical tool for exploring oscillation modes therein.
A systematic exploration of such higher dimensional settings would be 
a natural step for future investigations.




\end{document}